\documentclass[twocolumn,showpacs,amssymb,american,prl]{revtex4}
\usepackage{amsmath}
\usepackage{graphicx}
\usepackage{babel}

\begin{document}

\title{Full control by locally induced relaxation}

\author{Daniel Burgarth$^{1}$ and Vittorio Giovannetti$^{2}$}

\affiliation{$^{1}$Computer Science Department, ETH Z\"urich, CH-8092 Z\"urich,
Switzerland\\
$^{2}$NEST-CNR-INFM \& Scuola Normale Superiore, piazza dei Cavalieri
7, I-56126 Pisa, Italy}

\begin{abstract}
We demonstrate a scheme for controlling a large quantum system by
acting on a small subsystem only. The local control is mediated to
the larger system by some fixed coupling Hamiltonian. The scheme allows
to transfer \emph{arbitrary and unknown} quantum states from a memory
on the large system ({}``upload access'') as well as the inverse
({}``download access''). We study sufficient conditions of the coupling
Hamiltonian and give lower bounds on the fidelities for downloading and
uploading.
\end{abstract}


\maketitle

The unitarity of Quantum Mechanics implies that the information
lost by an open system during its dynamical evolution
must  be contained in the environment and, possibly,
in the correlations between the  system and the environment~\cite{Hayden2004}.
In the context of repetitive applications of the same quantum transformation 
this fact has been exploited to achieve noise protection~\cite{VIOLA3},
cooling, state preparation~\cite{
HOMOGENIZATION1,WELLENS}, 
and  quantum state transfer~\cite{MEMORYSWAP}. 
The mathematical
aspects related with  the convergence of the associated 
trajectories
are well studied 
(see Ref.~\cite{Gohm2004} and references therein).
It has however been largely overlooked that the relaxation effects 
associated with these approaches
 can also imply \emph{full control} by acting on a \emph{local} subsystem only.
This is analogous to what happens in the case of
the universal quantum interface of  
Ref.~\cite{SETH} where 
the local control  is mediated to the whole system by some fixed
coupling Hamiltonian.
Once this is achieved, 
apart from cooling and state preparation, 
it is also possible to perform
arbitrary quantum data processing (e.g. measurements, unitary rotations).

In this paper we discuss
an explicit protocol for universal control of a composite system
by operating on it with a simple repetitive local quantum transformation
and we provide lower bounds for fidelities obtainable after 
finitely many steps. 
Furthermore turning the problem into a
graph theoretic one we provide
an easy-to-check sufficient criterion to verify 
if a given global network Hamiltonian is capable of
mediating control. 

\begin{figure}
\begin{centering}\includegraphics[width=0.75\columnwidth]{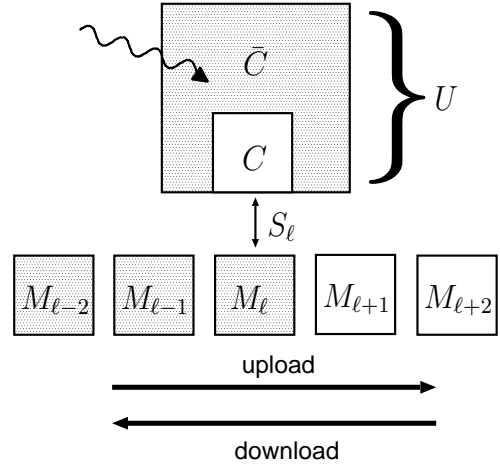}\par\end{centering}
\caption{\label{fig:memory}The systems $C$ and $\bar{C}$ are 
coupled through the time independent Hamiltonian $H$. 
The  system $C$ is 
controlled by performing regular swap operations $S_{\ell}$ between
it and a quantum memory $M_{\ell}.$ 
In CCD-like application $\bar{C}$ is an array of permanently coupled sensors
 that are used to probe external signals (schematically sketched by the
wiggling curve): information from the sensor is extracted 
through $C$.  
}
\end{figure}

The results presented here pave the way to 
new  applications of quantum control and quantum computation. Arguably 
our downloading and uploading 
protocols  (see below) may be useful for the
control of quantum hard drives and  
quantum RAM~\cite{QRAM}, or 
CCD-like application 
for the external control 
of permanently coupled arrays of sensors (see caption of Fig~\ref{fig:memory}). 
A large scale experimental realization of the
scheme discussed in this paper is not realistic at the moment
since it requires the ability of performing full quantum computation
on a large memory system. 
However proof-of-principle tests could be 
probably realized in hybrid quantum networks by 
exploiting the methods proposed in Ref.~\cite{TOMMASO} (e.g.   
controlling a permanently coupled optical lattice --
where local control is generally difficult -- by coupling it to a fully 
controllable array of 
trapped ions).  

\paragraph*{Protocol:--\label{sec:Protocol}}

We consider a tripartite finite dimensional Hilbert space
$\mathcal{H}=\mathcal{H}_{C}\otimes\mathcal{H}_{\bar{C}}\otimes\mathcal{H}_{M}.$
Full control (i.e. the ability to prepare states and apply
unitary transformations) is assumed on system $C$ and $M,$ but
no (direct) control is given on system $\bar{C}.$ System
$C$ and $\bar{C}$ are coupled by some time-independent Hamiltonian
$H.$ We will show  that under certain assumptions, if the system $C\bar{C}$
is initialized in an arbitrary state we can transfer (``download'')
this state into the system $M$ by applying some simple operations which
act locally on $M$ and $C$. Likewise, by initializing the system $M$ in the correct
state, we can ``upload''
arbitrary states on the system
$C\bar{C}$. These two schemes 
ensure full controllability of $C{\bar{C}}$: for instance  
one can perform 
arbitrary quantum operation on 
such system  by transferring its state into $M$, applying
the equivalent operation there, and transferring the resulting state back to $C{\bar{C}}$.
In this context $M$ functions as a 
\emph{quantum memory} and must be at least as
large as the system $C\bar{C}.$ As sketched in Fig.~\ref{fig:memory}
we can imagine it to be split into sectors $M_{\ell},$ 
$\mathcal{H}_{M}=\bigotimes_{\ell=1}^{L}\mathcal{H}_{M_{\ell}}$
with the $\mathcal{H}_{M_\ell}$ being isomorphic to $\mathcal{H}_C$ 
(i.e. $\textrm{dim}\mathcal{H}_{M_{\ell}}=\textrm{dim}\mathcal{H}_{C}$).

For downloading 
we assume that the memory $M$ is initialized in 
$|e\rangle_{M}\equiv\bigotimes_{\ell}|{e}\rangle_{M_{\ell}}$
with the vectors $|{e}\rangle$  to be defined in the following.
 To download an arbitrary initial state $|\psi\rangle_{C\bar{C}}$ 
of $C\bar{C}$ into the memory
$M$ we perform a sequence of unitary gates between $M$ and ${C}$,
intermitted by the time evolution $U=\exp[ -iHt ]$ on $C\bar{C}$ for
some fixed time interval $t$. More specifically,
at step $\ell$ of the protocol we perform a unitary swap $S_{\ell}$
between system $C$ and system $M_{\ell}$. The protocol stops after the $L$th swap
operation. The resulting global 
transformation  is thus represented
by the unitary operator \begin{equation}
W\equiv U S_{L}US_{L-1}\cdots U S_{\ell}\cdots U S_{1}. \label{WWWW}
\end{equation}
As we will see in the next section, the reduced evolution of
the system $\bar{C}$ under the protocol can be expressed in terms
of the completely positive trace preserving (CPT) map $\tau$ 
 defined in Eq.~(\ref{mapdef}).
Our main assumption is that the system $\bar{C}$ is \emph{relaxing}~\cite{TERHAL}
under repetitive application of $\tau,$ i.e. $\lim_{n\rightarrow\infty}\tau^{n}(\rho)=\rho_{*}$
for all initial states $\rho.$ This behavior is also called {\em mixing}~\cite{STRICTCONTRATIONS,Giovannetti}
or {\em absorbing}~\cite{Gohm2004}. 
In what follows we will focus on the case in which 
$\rho_{*}$ is a  pure state $|{E}\rangle_{\bar{C}}\langle {E}|$. 
When this happens 
it is possible to show
 that, 
for sufficiently large
$L$,  the  transfer of $|\psi\rangle$ 
from $C\bar{C}$ into $M$ can be done with arbitrarily high fidelity
~\cite{MEMORYSWAP} and the transformation which allows one
to recover  $|\psi\rangle$ from $M$ can be explicitly constructed.

For uploading
an arbitrary input state $|\psi\rangle$ from $M$ to $C\bar{C}$
one is tempted 
to {\em revert} the downloading protocol.
 Roughly speaking, the idea is to initialize the memory in the state that
it \emph{would have ended up in} after applying $W$ if system $C\bar{C}$
had started in the state we want to initialize. Then we apply the
\emph{inverse} of $W$ given by\begin{equation}
W^{\dag}=S_{1}U^\dag \cdots S_{\ell} U^{\dag}\cdots S_{L-1}U^{\dag}
S_{L} U^\dag \label{Wdag} \;.\end{equation}
We will see that indeed this induces a unitary coding on $M$
such that arbitrary and unknown states can be
initialized on $C{\bar{C}}$. The reader has probably noticed however that the
transformation~(\ref{Wdag})  is generally 
unphysical in the sense that it requires
backward time evolution of $C{\bar{C}}$, i.e. one has to wait \emph{negative} time
steps between the swaps. 
For this reason, even though the transformation originated from 
$W^\dag$ is coherently defined at a mathematical level, it
  cannot be considered as a proper uploading algorithm
for transferring states from $M$ to $C{\bar{C}}$: to stress this 
we will call the transformation associated to Eq.~(\ref{Wdag})
the {\em reverse-downloading} protocol.  
A proper 
uploading algorithm will be defined in the final part of the paper  
by imposing an extra hypothesis on the
$C{\bar{C}}$ couplings and by adopting a simple 
change of perspective. 
For the moment we neglect this point and simply focus  
on the convergence properties of the downloading and  the reverse-downloading 
algorithms associated with 
Eqs.~(\ref{WWWW}) and~(\ref{Wdag}).

\paragraph*{Cooling: --}
We start by showing that the action of $W$ on $C{\bar{C}}$ is 
effectively equivalent to a {\em cooling} process which transfers
any initial state into $|{e}\rangle_C|{E}\rangle_{\bar{C}}$.
Let $|\psi\rangle_{C\bar{C}}\in\mathcal{H}_{C\bar{C}}$
be an arbitrary state. We notice that the $C$ component of $W|\psi\rangle_{C\bar{C}}|{e}\rangle_{M}$
is always $|{e}\rangle_{C}$. 
Therefore we can write
\begin{eqnarray}
W|\psi\rangle_{C\bar{C}}|{e}\rangle_{M} \label{eq:main} =|{e}\rangle_{C}
\left[\sqrt{\eta}|{E}\rangle_{\bar{C}}|\phi\rangle_{M}+\sqrt{1-\eta}|\Delta\rangle_{\bar{C}M}\right] \end{eqnarray}
with $|\Delta\rangle_{\bar{C}M}$ being a normalized vector of $\bar{C}M$
which satisfies the identity 
${_{\bar{C}}\langle}{E}|\Delta\rangle_{\bar{C}M}=0$.
 It is worth stressing that the decomposition~(\ref{eq:main}) is unique and
that $\eta$, $|\phi\rangle_{M}$
and $|\Delta\rangle_{\bar{C}M}$ are typically complicated functions of 
the input state $|\psi\rangle_{C\bar{C}}$.
The quantity $\eta$ plays an important role: it gives us the fidelity
between  the initial state of $C{\bar{C}}$ and the target state 
 $|{e}\rangle_C|{E}\rangle_{\bar{C}}$ of the cooling process.
An expression for $\eta$ can be obtained
by focusing on the reduced density matrix of the subsystem $\bar{C}$.
From our definitions it follows that 
after the first step of the protocol
 this is 
\begin{eqnarray}
\tau(\rho_{\bar{C}}) & \equiv & 
\textrm{tr}_{CM}\left[US_{1}\left(|\psi\rangle_{\bar{C}C}\langle\psi|\otimes|{e}
\rangle_{M}\langle {e}|\right)S_{1}U^{\dag}\right]
\nonumber \\
&=&\textrm{tr}_{C}\left[U\left(\rho_{\bar{C}}\otimes|{e}\rangle_{C}\langle{e}
|\right)U^{\dag}\right]
\label{mapdef}
\;,\end{eqnarray}
 with $\rho_{\bar{C}}\equiv \textrm{tr}_{C}\left[|\psi\rangle_{\bar{C}C}\langle\psi|\right]$ being the reduced density matrix associated with the 
initial state $|\psi\rangle_{\bar{C}C}$.
Reiterating this expression we notice that
the state of $\bar{C}$ after $L$ steps can be obtained by
successive application of  the map (\ref{mapdef}). Consequently   
Eq.~(\ref{eq:main}) 
 gives
$\eta={}_{\bar{C}}\langle{E}
|\tau^{L}\left(\rho_{\bar{C}}\right)|{E}\rangle_{\bar{C}}$,
 which, according to the mixing properties of  $\tau$ given at the beginning
of the section, 
shows that $\eta\rightarrow1$ for
$L\rightarrow\infty$. Specifically we can use~\cite{TERHAL} to claim
that for all input states $|\psi\rangle$ the following inequality
holds 
\begin{eqnarray}
|\eta-1| 
\leqslant 
\|\tau^{L}\left(\rho_{\bar{C}}\right)-|{E}\rangle_{\bar{C}}\langle{E}|\|_{1}
\label{eq:fid11} 
\leqslant K \; \kappa^{L}\;L^{d_{\bar{C}}} \;,\end{eqnarray}
 where $K$ is a constant which depends upon $d_{\bar{C}}\equiv\mbox{dim}\mathcal{H}_{\bar{C}}$
and  $\kappa\in]0,1[$ is the second largest of the moduli of
eigenvalues of the map $\tau$.


\paragraph*{Coding transformation:--\label{sec:Coding-transformation}}

Let us now derive the decoding/encoding transformation that relates states
on the memory $M$ to the states of $C\bar{C}.$
The idea is to apply the decomposition~(\ref{eq:main}) to each element
of a given orthonormal basis $\left\{ |\psi_{k}\rangle_{C\bar{C}}\right\} $
of $\mathcal{H}_{C\bar{C}}$, 
and to define the linear operator $D$ on ${\cal H}_{M}$ which,
for all $k$, performs the
transformation \begin{eqnarray}
D|\psi_{k}\rangle_{M}=|\phi_{k}\rangle_{M}\label{DEFD}\;.\end{eqnarray}
In this expression $|\psi_{k}\rangle_{M}$ are orthonormal vectors of $M$ 
used to represent the states $|\psi_{k}\rangle_{C\bar{C}}$
of $\mathcal{H}_{C\bar{C}}$ on $M$
(formally they are 
obtained by a partial
isometry from $\bar{C}C$ to $M$).
The vectors $|\phi_{k}\rangle_{M}$ instead are connected to the
$|\psi_{k}\rangle_{C\bar{C}}$ through Eq.~(\ref{eq:main}).
Typically, for finite values of $L$,  the $|\phi_{k}\rangle_{M}$ will not
be orthogonal.  However 
it is possible to show that they become asymptotically orthogonal 
in the limit of $L\gg1$. 
To see this we use the unitarity of the transformation $W$ and the orthogonality
of $|\psi_k\rangle_{C\overline{C}}$. Indeed from Eq.~(\ref{eq:main})
one can easily verify the following identity
\begin{eqnarray}
\delta_{kk'} \label{QQQ}  &=& \sqrt{\eta_k \eta_{k'}} 
\; {_{M}\langle}\phi_{k}|\phi_{k'}\rangle_{M} \\ \nonumber 
&&+\sqrt{(1-\eta_{k})(1-\eta_{k'})}\;{_{\bar{C}M}\langle}
\Delta_{k}|\Delta_{k'}\rangle_{\bar{C}M}\;. \end{eqnarray}
Defining 
$\eta_{0}\equiv\min_{k}\eta_{k}$, 
 we notice that for 
sufficiently large $L$ this is a strictly positive quantity 
and converges to $1$ --- see Eq.~(\ref{eq:fid11}). 
From the identity~(\ref{QQQ}) it follows then that for $k\neq k'$
one can write 
$|_{M}\langle\phi_{k}|\phi_{k'}\rangle_{M}| \leqslant (1-\eta_0)/\eta_0$.
This  can now be used to bound the eigenvalues
$\lambda_{j}$ of the linear operator $D^{\dag}D$. Indeed 
the  Cauchy-Schwartz inequality yields 
$|\lambda_{j}-1|\leqslant  d_{C\bar{C}}\;({1-\eta_{0}})/{\eta_0}$,
with $d_{C\bar{C}}\equiv\dim\mathcal{H}_{C\bar{C}}$. Take now
a polar decomposition $D=PV$ with $P$ being positive semidefinite.
According to~\cite[p 432]{HORNJOHNSON}
$V$ is the \emph{best unitary
approximation} of $D$. In our case it satisfies the relations
\begin{eqnarray}
||D-V||_{2}^{2} & = & \sum_{j}\left[\sqrt{\lambda_{j}}-1\right]^{2}
 \leqslant  \sum_{k}\left|\lambda_{j}-1\right|\nonumber \\
 & \leqslant &  d_{C\bar{C}} (d_{C\bar{C}} -1)({1-\eta_{0}})/{\eta_0} \;.
\label{Q0}\end{eqnarray}
This is a key equation: 
thanks to Eq.~(\ref{eq:fid11}), it shows that $D$ can be
approximated arbitrary 
well by the unitary operator $V$ for $L\rightarrow\infty$.

\paragraph*{Fidelities:--}
In what follows we will use $V^{\dag}$ and $V$ as our downloading  and
reverse-downloading transformation, respectively. In particular, $V^{\dag}$ will
be used to recover the input state $|\psi\rangle_{C\bar{C}}$ of the
chain after we have (partially) transferred it into $M$ through the
unitary $W$ (i.e. we first act on $|\psi\rangle_{C\bar{C}}|{e}\rangle_{M}$
with $W$, and then we apply $V^{\dag}$ on $M$). Vice-versa, in
order to upload a state $|\psi\rangle_{M}$ on $C\bar{C}$ by using the
reverse-downloading protocol
we first prepare $C{\bar{C}}$ in
$|eE\rangle_{C\bar{C}}$~\cite{NOTE3},  then we apply to it
the unitary transformation $V$ and finally we apply $W^{\dag}$.

With these choices the fidelity for downloading
a state $|\psi\rangle_{M}$ is simply 
given by
$F_{d}(\psi)\equiv{_{M}\langle}\psi|V^{\dag}\; \rho_{M}\; 
V|\psi\rangle_{M} $
 where $\rho_{M}$ is the state of the memory after $W$, i.e.
\begin{eqnarray}
\rho_{M} & \equiv & \textrm{tr}_{C\bar{C}}\left[W(|\psi\rangle_{C\bar{C}}\langle\psi|\otimes|{e}\rangle_{M}\langle{e}|)W^{\dag}\right]\nonumber \\
 & = & \eta\;|\phi\rangle_{M}\langle\phi|+(1-\eta)\;\sigma_{M}\;,
\label{eq2000}
\end{eqnarray} 
(here we used Eq.~(\ref{eq:main}) 
and introduced the density matrix $\sigma_{M}\equiv \textrm{tr}_{\bar{C}}[
|\Delta\rangle_{\bar{C}M}\langle\Delta|]$). 
Analogously the fidelity for 
transferring a state $|\psi\rangle$ from $M$ to $C\bar{C}$
through the reverse-downloading protocol is given by 
\begin{eqnarray}
F_{rd}(\psi) &\equiv& \nonumber 
{_{C\bar{C}}\langle}\psi|\textrm{tr}_{M}\big[ 
W^{\dag}V\big(|\psi\rangle_{M}\langle\psi| \\ && 
\qquad \otimes|{eE}\rangle_{C\bar{C}}\langle {eE}|\big)
V^{\dag}W\big]|\psi\rangle_{C\bar{C}}.
\nonumber 
\end{eqnarray}
A bound for $F_{r}$ and $F_{rd}$ follows
by noticing that 
both these quantities satisfy the inequality
\begin{eqnarray}
F(\psi) 
 & \geqslant & \eta\;|{}_{M}\langle\phi|V|\psi\rangle_{M}|^{2}\;.\label{fin1}\end{eqnarray}
For $F_d$ this simply comes by expressing it in terms of Eq.~(\ref{eq2000})
and by neglecting a positive contribution proportional to $1-\eta$. 
For $F_{rd}$ instead  the inequality~(\ref{fin1}) follows by
 replacing the trace
over $M$ with the expectation value on $|0\rangle_{M}$ and by using 
Eq.~(\ref{eq:main}). One can  now estimate the scalar product on the right hand side
of Eq.~(\ref{fin1}) by observing that 
$|_{M}\langle\phi|V|\psi\rangle_{M}| 
\geqslant 
|_{M}\langle\phi|D|\psi\rangle_{M}|-|_{M}\langle\phi|D-V|\psi\rangle_{M}|$.
This can be further bounded by employing the inequality~(\ref{Q0}) 
and the fact that 
$|_{M}\langle\phi|D-V|\psi\rangle_{M}|\leqslant||D-V||_{2}$.
If $|\psi\rangle_{M}$ is a vector  of  the basis $|\psi_{k}\rangle_{M},$
then $|_{M}\langle\phi|D|\psi\rangle_{M}|=1$ by the definition -- see Eq.~(\ref{DEFD}). 
For \emph{generic} $|\psi\rangle_{M}$ instead  some simple algebra yields
$\sqrt{\eta}\;|_{M}\langle\phi|D|\psi\rangle_{M}|\;\geqslant\sqrt{\eta_{0}}\;-
\; d_{C\bar{C}}\;(1-\eta_{0})/\eta_0$. 
Replacing all this into Eq.~(\ref{fin1}) we finally get
\begin{eqnarray}
F&\geqslant& {\eta_{0}}\;-4\; d_{C\bar{C}}\;
\sqrt{(1-\eta_{0})/\eta_0}
\;,\label{fin1000}\end{eqnarray}
which holds for $F=F_r, F_{rd}$.
This  is a lower  bound for the fidelity of  the downloading 
and reverse-downloading protocols: 
it is probably not tight but it is sufficient to show that $F_r$ and $F_{rd}$ 
converge to $1$ in the limit of large $L$~\cite{NOTE1}.
According to Eq.~(\ref{eq:fid11}) such convergence is exponentially fast in $L$
even though, not surprisingly, the bound deteriorates as the size $d_{\bar{C}}$ 
of the controlled system increases.

\paragraph*{Uploading protocol:--}
Let us now come back to the question 
about the operation $W^{\dag}$ being unphysical. 
To define a proper uploading protocol 
consider a modified scenario in which $C{\bar{C}}$ is
replaced by a isomorphic system $C^\prime {\bar{C^\prime}}$  
characterized by the Hamiltonian ${H^\prime}=-H$. 
In this scenario
the downloading protocol is described by  the operator ${W^\prime} 
= {U^\prime} S_L {U^\prime} S_{L-1} \cdots {U^\prime}S_1$
with ${U^\prime} \equiv \exp[-i {H^\prime}t]$
while the corresponding reverse-downloading transformation by 
the operator 
$({W^\prime})^\dag = S_1 
({U^\prime})^\dag \cdots  S_{L-1} ({U^\prime})^\dag S_L  
({U^\prime})^\dag$.
Since $U^\prime = U^\dag$ it is not 
difficult to observe that 
the ``unphysical'' reverse-downloading algorithm of $C^\prime{\bar{C^\prime}}$ 
induces  a proper uploading transformation for   $C{\bar{C}}$.
To ensure that such algorithm converges it is hence sufficient to
study the downloading and reverse-downloading 
protocols associated with ${W^\prime}$ and ${W^\prime}^\dag$. 
According to our previous results 
this  can be  done by focusing on   the CPT map 
\begin{equation}
{\tau}^\prime(\rho_{\bar{C}})\equiv\textrm{tr}_{C}\left[U^{\dag}\left(\rho_{\bar{C}}\otimes|{e}\rangle_{C}\langle {e}|\right)U\right]\;, \end{equation}
which replaces $\tau$ of Eq.~(\ref{mapdef}).
It is then sufficient to assume  ${\tau^\prime}$ 
to be ergodic with pure fixed point~$|{E}\rangle_{\bar{C}}$.
When this happens 
we can 
define a transformation ${D^\prime}$ as in (\ref{DEFD}) and its
unitary part $V^\prime$. 
The latter is the coding transformation which  will be used  
for the uploading protocol of $C{\bar{C}}$.
Consequently the fidelity $F_{up}(\psi)$ associated with such algorithm 
is bounded as in Eq.~(\ref{fin1000}) with $\eta_0$ being lower bounded 
by Eq.~(\ref{eq:fid11}) where 
the parameters $K$ and $\kappa$ of $\tau$  have being
replaced by the corresponding quantities of $\tau^\prime$.
It should be noted that the definition of uploading protocol given here
is more general than in other schemes relying on time-reversal symmetries~\cite{WELLENS}.

\begin{figure}[t]
\begin{centering}\includegraphics[width=0.7\columnwidth]{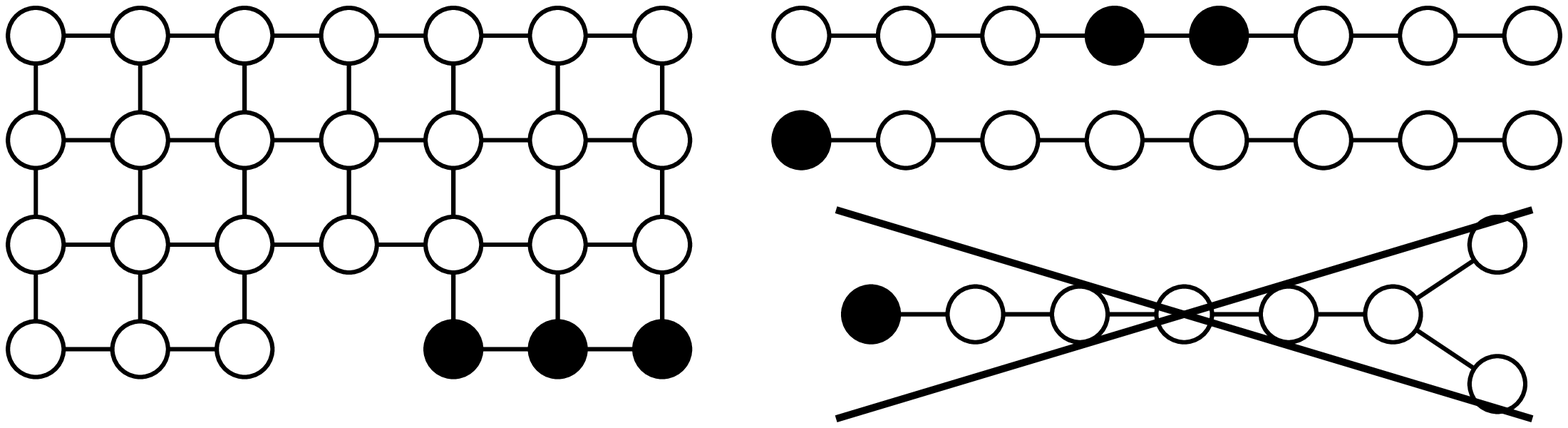}\par\end{centering}
\caption{\label{fig:graphs}Examples of graphs coupled by Heisenberg-like interaction
that can be controlled by acting on the black qubits only. The lower
right graph provides a counterexample.}
\end{figure}

\paragraph*{A condition for controllability:--}
The mixing properties of $\tau$ and $\tau^\prime$ are 
typically independent (see for instance Ref.~\cite{Gohm2004}): this
makes it difficult to give a general condition for  
the full controllability of $C{\bar{C}}$ (i.e.
convergence of both the downloading and uploading protocols). 
Notably however  
a generic statement can be made
using a result of Ref.~\cite{Giovannetti}. For the sake of simplicity here we
will focus on the case in which 
$C\bar{C}$ is a network 
of coupled spins 1/2 particles. 
According to~\cite{Giovannetti}
we have that {\em i)}  if the
Hamiltonian $H$ of $C{\bar{C}}$ 
preserves the number of spin excitations and {\em ii)} the vector 
$|{e}\rangle_{C}|E\rangle_{\bar{C}}$ is the only eigenstate 
with $C$ in $|{e}\rangle_C$
then the map $\tau$ is mixing with fix point $|{E}\rangle_{\bar{C}}$
(here $|{e}\rangle_C$ and $|{E}\rangle_{\bar{C}}$ represent 
states with all spins aligned down). 
However the Hamiltonian  $H^\prime=-H$
associated with  $\tau^\prime$ has the {\em same} eigenvectors of $H$: hence
the conditions {\em i)} and {\em ii)} 
also yields a sufficient criterion for determining  that 
$\tilde{\tau}$ is  mixing with fixed point~$|{E}\rangle_{\bar{C}}$.

The analysis further simplifies by 
 focusing on a two-sites interaction Hamiltonian. In this case   
a simple recursive analysis is sufficient to  check if 
$H$ satisfies the condition {\em ii)}.
For linear chains of spins this was discussed in 
Ref.~\cite{MEMORYSWAP}: here we generalize this argument to
arbitrary topology.
To do so, 
define the graph $\cal{G}$ with the spins of the network
 as vertices, and
the non-Ising components of $H$ as edges. 
Introduce also the following color code:
a black vertex corresponds to a spin in $|\!\downarrow\rangle$ while
a  white vertex corresponds 
to a generic spin configuration (i.e. not necessarily $|\!\downarrow\rangle$). 
Consider now the case in which the subset $C$ 
of the graph has all black vertexes. 
Our goal is to determine if such  configuration  
is compatible with being  a non trivial eigenstate of  
the network Hamiltonian (that is an eigenstate  in which
not all the vertices are black): if not, then  
the whole network can be controlled.
This can be checked by 
noticing that the excitations 
(i.e.  the white vertices) tend to 
propagate along the edges
when $H$ is applied to the graph: consequently  only certain 
distributions of  black and white  vertices are compatible with the eigenvector
structure of $H$ (they must allow 
certain interference effects that prevents the propagations
of the white vertices). To exploit  this property we introduce
the following
cellular graph automata: iff a vertex
is black and has exactly one white neighbor, then this neighbor will
turn black in the next step. Otherwise, vertices remain unchanged.
It follows that the final state associated with a generic
 initial configuration is the one which is compatible 
with being an eigenstate of $H$ and has the
minimum allowed number of black vertices.
Therefore to verify if the whole graph is controllable by operating on 
$C$ it is sufficient to initialize $C$ 
in the  all-black configuration and $\bar{C}$ in the
all-white configuration and let the graph evolve. 
If the final result has only black vertices  then we 
conclude that $C$ induces full control on $C{\bar{C}}$ 
(see Fig.~\ref{fig:graphs} for examples of controllable graphs). 
Note that this method allows us to say something about the structure
and the asymptotic dynamics of Hamiltonians that are in general very
far from analytic solvability, including disordered systems. 
\paragraph*{Conclusion:--}
We have shown that a for large class of physically realistic Hamiltonians, a Quantum Computer
can fully control a large system by inducing a relaxation on a small subsystem only.
The fidelity converges exponentially fast with the size of the memory, so the required overhead
of resources is low.
The results discussed here can be generalized to 
maps $\tau$ and $\tau^\prime$ are mixing with non-pure fixed point $\rho_*$.
In this case the fidelities of the protocols will be not necessarily
optimal. Still one can provide non trivial lower bound for this quantities
which depends upon the purity of $\rho_*$. DB acknowledges the Swiss National Science Foundation (SNSF) for
financial support.

\bibliographystyle{prsty}

\begin{thebibliography}{10}

\bibitem{Hayden2004}
P. Hayden \emph{et. al.}, Commun. Math. Phys. {\bf 250},
  371  (2004).

\bibitem{VIOLA3}
L. Viola,  S. Lloyd, and E. Knill,
Phys. Rev. Lett. {\bf 83},  4888  (1999).


\bibitem{HOMOGENIZATION1}
H.~V. Scarani {\it et~al.}, Phys. Rev. Lett. {\bf 88},  097905  (2002);
H. Nakazato, T. Takazawa, and K. Yuasa,  
{\emph ib.}
{\bf 90},  060401 (2003).

\bibitem{WELLENS}
T. Wellens  \emph{et. al.}, Phys. Rev. Lett.
  {\bf 85},  3361  (2000).

\bibitem{MEMORYSWAP}
V. Giovannetti and D. Burgarth, Phys. Rev. Lett. {\bf 96},  030501  (2006).

\bibitem{Gohm2004}
R. Gohm, {\em Noncommutative Stationary Processes} (Springer, New York, 2004).

\bibitem{SETH}S. Lloyd, A. Landahl, and J.J. Slotine,
Phys. Rev. A {\bf 69}, 012305 (2004).

\bibitem{QRAM}V. Giovannetti, S. Lloyd, and L. Maccone, in preparation.

\bibitem{TOMMASO}Z. Idziaszek,  T. Calarco, and P. Zoller,
Eprint quant-ph/arXiv:0704.1037v1.

\bibitem{STRICTCONTRATIONS}
M. Raginsky, Phys. Rev. A {\bf 65},  032306  (2002).

\bibitem{Giovannetti}
D. Burgarth and V. Giovannetti, Eprint quant-ph/0605197.

\bibitem{TERHAL}
B.~M. Terhal and D.~P. DiVincenzo, Phys. Rev. A {\bf 61},  022301  (2000).

\bibitem{NOTE3} Take this as an initial condition: in any case the system
can always be brought to $|eE\rangle_{C\bar{C}}$ by first cooling it.

\bibitem{HORNJOHNSON}
R.~A. Horn and C.~R. Johnson, {\em Matrix Analysis} (Cambridge University
  Press, Cambridge, U.K., 1990).

\bibitem{NOTE1}In particular the bound could be still optimized
by a proper choice of the basis $\{|\psi_k\rangle\}$ and of the time
interval~$t$.

\end{thebibliography}

\end{document}